\documentstyle[12pt,epsfig,rotating]{article}
\def\bd{
\begin{document}} \def\ed{\end{document}}
\def\bmp{\begin{minipage}} \def\emp{\end{minipage}}
\def\bcc{\begin{center}} \def\ecc{\end{center}}     \def\npg{\newpage}
\def\beq{\begin{equation}} \def\eeq{\end{equation}} \def\hph{\hphantom}
\def\be{\begin{equation}} \def\ee{\end{equation}} \def\r#1{$^{[#1]}$}
\def\n{\noindent} \def\ni{\noindent} \def\pa{\parindent}
\def\hs{\hskip} \def\vs{\vskip} \def\hf{\hfill} \def\ej{\vfill\eject}
\def\cl{\centerline} \def\ob{\obeylines}  \def\ls{\leftskip}
\def\underbar#1{$\setbox0=\hbox{#1} \dp0=1.5pt \mathsurround=0pt
   \underline{\box0}$}   \def\ub{\underbar}    \def\ul{\underline}
\def\f{\left} \def\g{\right} \def\e{{\rm e}} \def\o{\over} \def\d{{\rm d}}
\def\vf{\varphi} \def\pl{\partial} \def\cov{{\rm cov}} \def\ch{{\rm ch}}
\def\la{\langle} \def\ra{\rangle} \def\EE{e$^+$e$^-$} \def\pt{p_{\rm t}}
\def\bitz{\begin{itemize}} \def\eitz{\end{itemize}}   \def\vf{\varphi}
\def\btbl{\begin{tabular}} \def\etbl{\end{tabular}}   \def\kt{k_{\rm t}}
\def\btbb{\begin{tabbing}} \def\etbb{\end{tabbing}}
\def\beqar{\begin{eqnarray}} \def\eeqar{\end{eqnarray}}
\def\\{\hfill\break} \def\dit{\item{-}} \def\i{\item}
\def\bbb{\begin{thebibliography}{9} \itemsep=-1mm}
\def\ebb{\end{thebibliography}} \def\bb{\bibitem} \def\yct{y_{\rm cut}}
\def\bpic{\begin{picture}(260,240)} \def\epic{\end{picture}}
\def\akgt{\noindent{Acknowledgements}} \def\pt{p_{\rm t}} 
\def\fgn{\noindent{\bf\large\bf Figure captions}} \def\EE{e$^+$e$^-$\ }
\bd

\centerline{\Large A MONTE CARLO STUDY ON THE}
\vskip0.3cm

\centerline{\Large DYNAMICAL FLUCTUATIONS INSIDE}
\vskip0.3cm

\centerline{\Large QUARK AND ANTIQUARK JETS\footnote{Supported in part by the
National Natural Science Foundation of China under Grant No.19775018}}

\vskip 1.2cm

\cl {Zhang Kunshi\footnote{Permanent address: Department of Physics, Jingzhou 
Teacher's Collage 434100.}, Chen Gang and Liu Lianshou\footnote{E-Mail: 
liuls@iopp.ccnu.edu.cn}}
\vskip0.5cm

\cl{Institute of Particle Physics, Huazhong Normal University, Wuhan, China.}
\vskip 1.2cm

\begin{center}
{\large ABSTRACT}
\vskip 0.6cm
\begin{minipage}{14cm}
\hskip1cm
      The dynamical fluctuations inside the quark and antiquark 
      jets are studied using Monte Carlo method.
      Quark and antiquark jets are identified from the 2-jet events
      in \EE collisions at 91.2 GeV by checking them at parton level.
      It is found that
      transition point exists inside both of these two kinds of jets. 
      At this point the jets are circular in the transverse plane
      with respect to the property of dynamical fluctuations. 
      The results are consistent with the fact that 
      the third jet (gluon jet) was historically first discovered in 
      \EE collisions in the energy region 17-30  GeV.

\vskip2cm

Keywords: Quark jet, Antiquark jet, Dynamical fluctuations 

\vskip0.5cm

PACS: 13.85.Hd
\end{minipage}

\end{center}
\newpage
\section* {1   Introduction}

As is well known, in high energy collisions the hadronization of 
partons, being a soft process, can not be analyzed 
using perturbative QCD (pQCD). Therefore, phenomenological analysis 
has to be applied to study the properties of the final state hadronic system, 
in order to get information which would lead to a better understanding of the 
dynamics of strong interaction. Investigating the dynamical fluctuations 
inside jets produced by the fragmentation of partons is an effective way 
to explore the dynamics in high energy collisions. 

In a 2-jet event in \EE collisions, the two produced jets are expected to
be back to back and the thrust (or sphericity) axis coincides with the 
jet axes. Furthermore, the two jets being produced from a quark and an
antiquark respectively are expected to have similar property. Under these 
assumptions the study of the dynamical fluctuations inside a single jet 
could be performed through the investigation of the fluctuations in
a 2-jet event. Recently, this kind of investigation has been carried out 
both theoretically~\cite{LCHPrD} and  experimentally~\cite{CGDatong}.

In view of the importance of this problem, it is worthwhile studying the
dynamical fluctuations inside a single jet directly~\cite{ZKSHEP}. The
aim of this paper is to carry on this study for quark and antiquark jets 
separately using Monte Carlo method. Through this study, the influence 
of the above mentioned two assumptions are gotten rid of and the dynamical
fluctuation properties in the hadronization process of quark and antiquark
can be directly obtained. 

\section*{2 Selection of quark and antiquark jets}

Event samples of \EE collision at c.m. energy $\sqrt s=91.2$ GeV are
produced using LUND JETSET7.4 Monte Carlo code. Only charged particles 
are used in the calculation.

Two-jet events are selected using the Durham jet algorithm\cite{Durham}. 
In this scheme, a jet resolution variable $\yct$ is defined 
for every pair of particles (or jets) $i$ and $j$ in an event by: 
\beq
 y_{ij}=\frac{2\min(E_i^2,E_j^2)}{s}(1-\cos \theta_{ij})
\eeq 
where $E_i$ and $E_j$ are the energies of the two particles 
(or jets), $\theta_{ij}$ is the angle between them and  
$s$ is the square of c.m. energy of the event. 
Jets or particles with $y_{ij}\leq \yct$
are combined into a single jet. This procedure is 
repeated untill all pairs $i$ and $j$ satisfy $y_{ij}>\yct$. The 2-jet
events remained at the end of this process consist 
the 2-jet sample corresponding to the cut parameter $\yct$.

Since we are using Monte Carlo method, it is possible to study the
2-jet events both at parton level and at hadron level.
Quark jet and antiquark jet can then be identified through matching
the jets in these two levels using the following technique~\cite{YML}.

Firstly, identify the type of a single jet at parton level by 
checking every parton involved in this jet and giving a weight +1 to 
quark, -1 to antiquark and 0 to gluon. The sum of the weights
is expected to yield +1 for quark jet and -1 for antiquark jet. 
The cases where the sum of weights in a jet is neither +1 nor -1, or
the weights in the two jets of one event are both +1 or both -1 are
neglected. This cut condition throw away about 1\% of the total events.

Secondly, consider the configuration of the 2-jet event at parton and 
hadron levels.  The directions of hadron jets might not be exactly that
of the corresponding parton ones, since the effects of hadronization 
may change their directions. 

The type of a hadron jet can be determined by checking the parton jet
closest to it. The hadron jet closest to the parton level quark jet is 
considered as the quark jet and that closest to the parton level antiquark 
jet is identified as the antiquark jet. In this way, 
the quark jet subsample and the antiquark jet subsample are obtained.

\section*{3 Single-jet coordinate system }

   In the 2-jet events, the thrust (or sphericity) axis is supposed to be 
the direction in which the quark-antiquark move back to back with high 
momenta. Therefore, it is chosen as the longitudinal direction in the
physics analysis.  While when the single quark or antiquark 
jet is considered separately, the thrust axis is no longer
appropriate to be considered as the longitudinal direction. The direction 
of the total momentum of the particles inside the jet should be taken as 
the longitudinal axis instead. Thus a jet coordinate system with the third
axis pointing along the direction of jet momentum should be built up. 
The transformation from the lab system to the jet system is acomplished
through the following two steps. First, rotate the lab frame $o$-$xyz$ an angle 
$\varphi$ around the $z$ axis to form the
$o$-$x'y'z'$ frame and then turn the latter an angle $\theta$ around $y'$ axis 
to get the single-jet coordinate system $o$-$x''y''z''$, cf. Fig.1.  
This transformation can be expressed as:
\beq  
\f(\matrix{x'' \cr y'' \cr z'' \cr}\g)=
\f(\matrix{\cos \theta \cos \vf & \cos\theta\sin\vf & -\sin\theta\cr
-\sin\vf & \cos\vf & 0 \cr
\sin\theta \cos\vf & \sin\theta\sin\vf & \cos\theta \cr}\g)
\f(\matrix{x \cr y \cr z \cr}\g)
\eeq

\begin{picture}(180,250)
\put(48,115)
{
{\epsfig{file=fig1a.epsi,width=125pt,height=120pt}}
{\epsfig{file=fig1b.epsi,width=125pt,height=120pt}}
}
\end{picture}
\vskip -4cm

{\small { Fig.1 The transition from laboratory frame to  
              quark (antiquark) jet frame  }} 

\begin{picture}(180,250)
\put(9,-40)
{
{\epsfig{file=fig2.epsi,width=335pt,height=300pt}}
}
\end{picture}

\vskip -4cm

{\small {Fig.2 The distribution of the angle between the quark 
(antiquark) jet axis  \\
 \hskip 5.8cm and the thrust axis. (a) quark jet  (b) antiquark jet}} 

\vskip5mm

The distribution of the angular difference between the direction of
the quark (antiquark) jet axis and the thrust axis is shown in Fig.2. 
It can be seen from the figure that the jet axis is very close to 
the thrust axis.

\section*{4 Dynamical fluctuations}

The dynamical fluctuations are characterized by the anomalous scaling of
normalized factorial moments (NFM)\cite{BP} 
\begin{equation}
F_{q}(M)=\frac{1}{M}\sum\limits^M_{m=1}
\frac{\la n_m(n_m-1)\cdots(n_m-q+1)\ra}
{\la n_m\ra^q} \sim M^{\phi_q} \quad M\to \infty,
\end{equation}
where a region $\Delta$ in one-, two-, or three-dimensional phase space 
is divided into $M$ cells, $n_m$ is the multiplicity in the $m$th cell, 
and $\la\cdots\ra$ indicates vertically averaging over the event sample, 
consisting of quark or antiquark jets.

As Ochs pointed out, the dynamical fluctuations occured in 
higher-dimentional (2D or 3D) phase
space have the projection effect\cite{projection}
on the fluctuations in lower-dimentional space causing the second-order 1D NFM 
goes to saturation by the rule~\cite{ZGKX}: 
   \begin{equation}
       F_{2}^{(a)}(M_a)=A_a-B_aM_a^{-\gamma_a},\  \  \  \  \  \  \ 
       a=y, \pt, \varphi
   \end{equation} 
where the exponent $\gamma_a$ describes the rate of going to saturation 
of the NFM in the direction $a$ and is the most important characteristic 
for the higher-dimensional dynamical fluctuations. 
If the values of $\gamma$ in 
two directions are equal, $\gamma_a=\gamma_b$, the fluctuations are 
isotropic in the $a,b$ plane; while when $\gamma_a\not=\gamma_b$ 
the fluctuations are anisotropic in this plane~\cite{FFLPrl}. 
The nature of the dynamical fluctuations (or that of the 
fractal) can be expressed in terms of the Hurst 
exponent $H_{ab}$, which can be obtained from the values of $\gamma_a$
and $\gamma_b$ as~\cite{ZGKX} 
\begin{equation}
   H_{ab}=\frac{1+\gamma_b}{1+\gamma_a}
\end{equation}
   The dynamical fluctuations are isotropic (self-similar fractal) when 
$H_{ab}=1$, and anisotropic (self-affine fractal) when 
$H_{ab}\not=1$~\cite{FFLPrl}.

In our Monte Carlo simulation a total number of 2000000 events of 
\EE collisions at 91.2 GeV are produced by JETSET 7.4 generator. 
Different $\yct$ are used to select the 2-jet events in the 
phase spase region $[0<y<5]$, $[0.1<\pt< 3.0]$, $[0<\varphi<2\pi]$ and 
the second order 1D factorial moments $F_{2}(y)$, $F_{2}(\pt)$, 
$F_{2}(\varphi)$ are calculated respectively for both quark and 
antiquark jets. In order to avoid the 
influence of a non-flat distribution of the  
variables $y, \pt$ and $\varphi$ on the investigation 
of the dynamical fluctuations, all variables are transformed into their 
corresponding cumulant forms~\cite{cummulant}. 
\beq  
x(y)={{\int_{y_a}^{y} \rho (y)dy}\over {\int_{y_a}^{y_b} \rho (y)dy}} ,
      \qquad  x(\pt)={{\int_{p_{\rm ta}}^{\pt} \rho (\pt)\d\pt}\over
         {\int_{p_{\rm ta}}^{p_{\rm tb}} \rho (\pt)\d\pt}},
      \qquad  x(\varphi)={{\int_{\varphi_a}^{\varphi} \rho
     (\varphi)d\varphi}\over {\int_{\varphi_a}^{\varphi_b}
        \rho (\varphi)d\varphi}}.
\eeq

Then the fitting to the saturated curves for these second order NFM's vs. 
the partition number $M=1, 2, \dots, 40$ 
in 1D phase space is carried out and the corresponding parameters 
$\gamma_y, \gamma_{\pt}$ and $\gamma_{\varphi}$ are obtained. 
In order to eliminate the influence of the momentum conservation, the first 
(for $F_2(\vf)$ the first three) point(s) are omitted when fitting the 
data to Eq.(4). 

\begin{picture}(180,250)
\put(-105,-45)
{
{\epsfig{file=fig3.epsi,width=480pt,height=300pt}}
}
\end{picture}
\vskip1.0cm
\hskip2.5cm ($a$) \hskip6.6cm ($b$)
\vskip0.4cm
{\small {Fig.3 The parameter $\gamma$ as function of $\yct$. ($a$) quark jets 
($b$) antiquark jets  }}

\section*{5. Results and discusion}
  
The variation of the three $\gamma_a$ ($a=y,\pt,\vf$) 
with $\yct$ are shown In Fig.3.
From the figure some properties of the dynamical fluctuations inside 
the quark and antiquark jets can be 
extracted.

The most striking feature in Fig.3 is the variation of $\gamma_{\varphi}$ 
with the parameter $\yct$. At small values of $\yct$,
$\gamma_{\varphi}$ lies at the bottom. When $\yct$ 
increases $\gamma_{\varphi}$ increases rapidly and at a certain point 
$\gamma_{\varphi}$ crosses over $\gamma_{\pt}$, turning from 
$\gamma_{\varphi}<\gamma_{\pt}$ to $\gamma_{\varphi}>\gamma_{\pt}$.
A distinct point called transition point~\cite{LCHPrD}, 
where $\gamma_{\varphi}=\gamma_{\pt}$, exists in 
both quark- (Fig.3$a$) and antiquark (Fig.3$b$) jets with only slightly 
different values of $\yct$ --- 
0.0047 for quark jets and 0.0050 for antiquark jets. 

At the transition point, the Hurst exponents are
$$H_{y\pt}=\frac {1+\gamma_{\pt}}{1+\gamma_{y}}=0.70\pm0.03  
; \ \ 0.74\pm0.03 $$ 
$$H_{y\varphi}=\frac{1+\gamma_{\varphi}}{1+\gamma_{y}}=0.70\pm0.06 
; \ \ 0.75\pm0.04 $$
$$H_{\pt\varphi}=\frac{1+\gamma_{\varphi}}{1+\gamma_{\pt}}=1.00\pm0.10 
; \ \ 1.01\pm0.06 $$ 
The first values are for quark jets and the second ones for antiquark jets.  
These results indicate that at the transition point the dynamical 
fluctuations inside quark (antiquark) jet  
are anisotropic in the longitudinal-transverse planes, 
($y, \pt)$ and ($y, \varphi$), and isotropic in the transverse plane 
($\pt, \varphi$). This means that the quark and antiquark jets
are circular in the transverse plane with respect to the dynamical 
fluctuations at the transition point and therefore they are called 
circular jets~\cite{LCHPrD}. 
For these jets $H_{y\pt}\not=1$, $H_{y\varphi}\not=1$ showing that
the quark (antiquark) jet system is self-affine fractal\cite{FFLPrl} 

In the Durham algorithm used in our calculation the relative
transverse momentum $\kt$ is related to $\yct$ as~\cite{kt} 
\begin{equation}
\kt=\sqrt{\yct}\cdot\sqrt{s}
\end{equation} 
The value of $\kt$ corresponding to the transition point is 
approximately 6.25 GeV for quark jet and 6.46 GeV for antiquark jet.  
The values of various parameters for both quark and antiquark jets 
at the transition point are listed in Table I

\begin{center}
Table I Parameters $\gamma$, Hurst exponent $H$, cut-parameter $\yct$
and relative transverse momentum $\kt$ at the transition point \\

\vskip 0.5cm
{\small\center
\begin{tabular}{c|c|c|c|c|c|c|c|c}\hline
subsample & $\yct$ & $\gamma_{y}$  &$\gamma_{\pt}$  &$\gamma_{\varphi}$ &$H_{y\
pt} $&$H_{y\varphi}$&$H_{\pt\varphi}$ &$\kt$ (GeV)\\
  \hline
quark&$ 0.0047$&$1.464$&$0.730$&$0.735$
&$0.70$&$0.70$&$1.00$&$6.25 $\\
jet&$ \pm 0.0003$&$\pm 0.036$&$\pm 0.041$&$\pm 0.123$&$\pm 0.03$
&$\pm 0.06$&$\pm 0.10$&
$\pm 0.17$\\\hline
antiquark&$ 0.0050$&$1.409$&$0.791$&$0.803$
&$0.74$&$0.75$&$1.01$&$6.46 $\\
jet&$\pm 0.0009$&$\pm 0.033$&$\pm 0.039$&$\pm 0.059$&$\pm 0.03$
&$\pm 0.04$&$\pm 0.06$&
$\pm 0.58$\\\hline
\end{tabular}}\vskip 0.5cm
\end{center}

Having analyzed the dynamical fluctuations inside quark (antiquark) jets,   
it is interesting to discuss the following question.  
In high energy experiments some algorithm, e.g. the Durham 
algorithm, is used to select jets. In doing so, the jet sample obtained 
strongly depends on the cut parameter $\yct$.  
When the value of $\yct$ is chosen appropriately,
the selected jet will be directly observable in the experiments as a
bunch of particles. These directly observable jets are refered to as
visible jets~\cite{LCHPrD}. The question is: Is
the circular jets determined by the properties of the
dynamical fluctuations at the transition point consistent with 
the visible jets in experiments?

In order to see the relation 
between the circular jets determined by the transition point and the visible 
jets observed in experiments, let us notice the fact that it was in the energy 
region 17--30 GeV that a third jet (the gluon jet) 
was historically first observed in \EE collisions~\cite{3jet}. 
In Fig.4's are shown the ratios $R_3$ 
of 3-jet events as function of the relative transverse momentum 
$\kt$ at 5 different energies ranging from 11 to 91.2  GeV. The dashed 
vertical lines correspond to the transition point $\kt$=6.25 GeV and 
$\kt$= 6.46 GeV, respectively. 

\begin{picture}(180,250)
\put(39,40)
{
{\epsfig{file=fig4.epsi,width=350pt,height=180pt}}
}
\end{picture}
\vskip-1cm

\centerline
{\small {Fig.4  The ratio of $R_3$ as function of the relative transverse 
momentum $\kt$  }}

\vskip0.8cm

It can be seen from the figure that for the circular jets defined by 
the transition point $\kt=6.25\sim6.46$ GeV (dashed lines in the figure),
3-jet events become noticeable just at $\sqrt s \approx 17$ GeV. This means
that the third (gluon) jet first observed at $\sqrt s = 17$ GeV is 
consistent with being the circular jet. 

\section*{4. Conclusion}

In this paper we identify the hadronic quark- and antiquark-jets 
through matching them with the parton-level jets. 
Then we study the dynamical fluctuations 
inside the quark jets and the antiquark jets respectively using  Monte Carlo
simulation.  

The second order 1D factorial Moments of the three variables $y, \pt, \vf$ 
are calculated. By fitting the results to the projection formula Eq.(4)
the variation of the saturation exponents $\gamma_y, \gamma_{\pt}$ and 
$\gamma_{\varphi}$ with $\yct$ is obtained. A transition point 
exists in both the quark and the antiquark jets. At the transition point 
the dynamical fluctuations inside the quark jets and the antiquark jets, being  
similar to each other, are anisotropic in the longitudinal-transverse planes 
and isotropic in the transverse plane. These jets are refered to as
circular jets. 

The percentage $R_3$ of 3-jet events for the jets of different $\kt$ have 
been calculated. It turns out that for the circular jets defined by 
the transition point $\kt=6.25\sim6.46$ GeV, the percentage $R_3$ of 
3-jet events becomes noticeable just at $\sqrt s \approx 17$ GeV, where
the 3-jet event was historically first observed.
Therefore, the circular jets defined by the $\kt$ at transition point
is consistent to be the visible jets, which are directly observable in
experiments.
 
\vskip 0.6cm
\subsection*{Acknowledgement}
The authors thank Yu Meiling for valuable discussions.

\vskip 1.5cm

\def\J#1#2#3#4{{\it #1} {\bf #2}, {#3} (#4)}
\def\PRL{Phys. Rev. Lett.} \def\PRep{Phys. Rep.}
\def\PR{Phys. Rev.}   \def\ZP{Z Phys.}
\def\NP{Nucl. Phys.}  \def\PL{Phys. Lett.}

\ed